\documentclass[RNAAS]{aastex62}
\shorttitle{}
\shortauthors{Zhang}

\begin{document}


\title{The IceCube coincident neutrino event is unlikely to be physically associated with LIGO/Virgo S190728q}


\author{Bing Zhang}
\affil{Department of Physics and Astronomy, University of Nevada Las Vegas, NV 89154, USA}
\email{zhang@physics.unlv.edu}








\section*{}

On 2019 July 28th, the LIGO Scientific Collaboration and the Virgo Collaboration reported the detection of a compact binary merger gravitational wave (GW) event LIGO/Virgo S190728q \citep{LIGO1,LIGO2}. The false alarm rate was $2.5\times 10^{-23}$ Hz. According to the second circular \citep{LIGO2}, the classification of the GW event, in order of descending probability, is BBH (95\%), MassGap (5\%), NSBH ($<$1\%), or BNS ($<$1\%). The  luminosity distance is $d_{\rm L} = 874 \pm 171$ Mpc.

The IceCube team first reported an upper limit on the neutrino flux from the GW event \citep{IceCube0}, but reported \citep{IceCube1,IceCube2,IceCube3} a track-like muon neutrino event in spatial and temporal coincidence with LIGO/Virgo S190728q shortly afterwards. The time offset is $-360$ s with respect to the GW event trigger. The p-values are 0.014 ($2.21 \sigma$) and 0.010 ($2.33 \sigma$) for generic transient search and Bayesian search, respectively \citep{IceCube3}.

These p-values are not small enough to allow a claim on the physical association between the IceCube event and S190728q. On the other hand, since a true GW-neutrino association would have profound implications, it is interesting to assess the physical plausibility of the association between the two multi-messenger events. We show below that a physical association is essentially impossible based on an energy budget argument.

Let us assume that the track-like neutrino event detected by IceCube is indeed from S190728q. One can estimate the neutrino fluence as ${\cal F}_\nu = \epsilon_\nu / A_{\rm eff} = 1.6 \times 10^{-5} \ {\rm erg \ cm^{-2}} (\epsilon_\nu/100 \ {\rm TeV}) (A_{\rm eff} / 10^3 {\rm m}^2)^{-1}$, where $\epsilon_\nu$ (normalized to 100 TeV) is the energy of the neutrino (not reported), and $A_{\rm eff}$ (normalized to $10^3 \ {\rm m^2}$) is the effective area of the IceCube detector \citep{IceCube}. For the Planck 2015 cosmological parameters \citep{planck}, the corresponding redshift for $d_{\rm L} = 874 \pm 171$ Mpc is $z = 0.179 \pm 0.032$. The total isotropic neutrino emission energy from the source is $E_{\rm \nu,iso} = 4 \pi d_{\rm L}^2 (1+z)^{-1} {\cal F}_\nu = (1.24^{+0.49}_{-0.42} \times 10^{51} \ {\rm erg}) \ (\epsilon_\nu/100 \ {\rm TeV}) (A_{\rm eff} / 10^3 {\rm m}^2)^{-1}$. This should be the lower limit of the total energy that is dissipated at the source to power the neutrino emission. If the neutrino emission is beamed, the energy budget is smaller by a factor of $f_b \equiv \Delta \Omega/4\pi$, where $\Delta \Omega$ is the beaming angle of the neutrino emission.

One obvious way to make bright neutrino emission is to assume that one of the merger members is a neutron star (even if the final NSBH probability is $<$1\%). At $\sim 360$ s {\em before} the GW event trigger, the NS is not tidally disrupted. The most plausible energy source from an NS is the magnetic energy of its magnetosphere. Indeed, interactions of the magnetospheres of two NSs have been invoked as the main mechanism to power precursor radiation for NS-NS mergers \citep[e.g.][]{hanson01,piro12,wang16}. The maximum energy that can be tapped via magnetospheric interactions is the total magnetic energy of the entire NS magnetosphere, which is $E_{\rm B} = (1/6) B^2 R^3 = 1.67 \times 10^{47} \ {\rm erg} \ B_{15}^2 R_6^3$ \citep{zhang14}, where $B$ (normalized to $10^{15}$ G for the most magnetized NSs) is the surface magnetic field and $R$ (normalized to $10^6$ cm) is the radius of the NS. One can see that $E_{\rm B}$ is smaller than $E_{\rm \nu,iso}$ by four orders of magnitude. Since there is no bright short gamma-ray burst associated with S190728q \citep{Fermi}, the viewing direction is not in a narrow jet. It is impossible to adjust $f_b$ to be small enough to account for this discrepancy. 

Some special progenitor models invoking two BHs inside a massive star \citep[e.g.][]{loeb16,janiuk17} may meet the energy budget requirement. In principle, one may fine tune the jet launching time to be before the merger time \citep[e.g.][]{dorazio18}. However, even for these scenarios, one would expect that the jet power is stronger after the merger, so that the neutrino emission luminosity should be higher around or after the merger time than at $\sim 360$ s before the merger.

The argument discussed here applies to all future putative GW-neutrino associations. Neutrino events after a GW merger signal would be much more credible than before the merger signal for physical associations.

This argument has applied the assumption that both GWs and neutrinos travel with or very close to the speed of light, so that there is no additional fundamental-physics-related time offset between the two signals.



\end{document}